\documentclass[aoas,MSNbibl,nameyear,seceqn,dvips]{arximspdf}
\usepackage{mathbh,dcolumn,url,breakurl}
\usepackage{graphicx}

\doi{10.1214/14-AOAS739} 
\volume{8}
\issue{2}
\pubyear{2014}
\firstpage{1256}
\lastpage{1280}

\makeatletter
\newcolumntype{d}[1]{D{.}{.}{#1}}
\newproclaim{mydef}{Definition}
\newcommand{\R}{\mathbb{R}}
\renewcommand{\P}{\mathbb{P}}
\newcommand{\E}{\mathbb{E}}
\newcommand{\logit}{\operatorname{logit}}

\def\argmax{\mathop{\operatorname{arg\,max}}}
\def\argmin{\mathop{\operatorname{arg\,min}}}
\def\mathbbm{\mathbh}
\makeatother

\begin{document}
\begin{frontmatter}

\title{Probability aggregation in time-series:
Dynamic hierarchical modeling of sparse expert beliefs\thanksref{T1}}
\runtitle{Probability aggregation in time-series}
\thankstext{T1}{Supported by a research contract to the University
of Pennsylvania and the University of California from the Intelligence
Advanced Research Projects Activity
(IARPA) via the Department of Interior National Business Center
contract number D11PC20061.}

\begin{aug}
\author[a]{\fnms{Ville A.}~\snm{Satop\"a\"a}\corref{}\ead[label=e1]{satopaa@wharton.upenn.edu}},
\author[a]{\fnms{Shane T.}~\snm{Jensen}\ead[label=e3]{stjensen@wharton.upenn.edu}},
\author[b]{\fnms{Barbara~A.}~\snm{Mellers}\ead[label=e4]{mellers@wharton.upenn.edu}},
\author[b]{\fnms{Philip~E.}~\snm{Tetlock}\ead[label=e5]{tetlock@wharton.upenn.edu}}
\and
\author[c]{\fnms{Lyle~H.}~\snm{Ungar}\ead[label=e2]{ungar@cis.upenn.edu}}
\runauthor{V.~A. Satop\"a\"a et al.}
\affiliation{University of Pennsylvania}
\address[a]{V.~A. Satop\"a\"a\\
S.~T. Jensen\\
Department of Statistics\\
The Wharton School\\
University of Pennsylvania\\
Philadelphia, Pennsylvania 19104-6340\\
USA\\
\printead{e1}\\
\phantom{E-mail:\ }\printead*{e3}}
\address[b]{B.~A. Mellers\\
P.~E. Tetlock\\
Department of Psychology\\
University of Pennsylvania\\
Philadelphia, Pennsylvania 19104-6340\\
USA\\
\printead{e4}\\
\phantom{E-mail:\ }\printead*{e5}}
\address[c]{L.~H. Ungar\\
Department of Computer\\
\quad and Information Science\\
University of Pennsylvania\\
Philadelphia, Pennsylvania 19104-6309\\
USA\\
\printead{e2}}
\end{aug}
%

\received{\smonth{9} \syear{2013}}
\revised{\smonth{3} \syear{2014}}

%
\begin{abstract}
Most subjective probability aggregation procedures use a single
probability judgment from each expert, even though it is common for
experts studying real problems to update their probability estimates
over time. This paper advances into unexplored areas of probability
aggregation by considering a dynamic context in which experts can
update their beliefs at random intervals. The updates occur very
infrequently, resulting in a sparse data set that cannot be modeled by
standard time-series procedures. In response to the lack of appropriate
methodology, this paper presents a hierarchical model that takes into
account the expert's level of self-reported expertise and produces
aggregate probabilities that are sharp and well calibrated both in- and
out-of-sample. The model is demonstrated on a real-world data set that
includes over 2300 experts making multiple probability forecasts over
two years on different subsets of 166 international political events.
\end{abstract}

%

\end{frontmatter}

\section{Introduction}
\label{intro}
Experts' probability assessments are often evaluated on \textit
{calibration}, which measures how closely the frequency of event
occurrence agrees with the assigned probabilities. For instance,
consider all events that an expert believes to occur with a 60\%
probability. If the expert is well calibrated, 60\% of these events
will actually end up occurring. Even though several experiments have
shown that experts are often poorly calibrated [see, e.g., \citet
{cooke1991experts,shlyakhter1994quantifying}], these are noteworthy
exceptions. In particular, \citet{wright1994coherence} argue that
higher self-reported expertise can be associated with better calibration.

Calibration by itself, however, is not sufficient for useful
probability estimation. Consider a relatively stationary process, such
as rain on different days in a given geographic region, where the
observed frequency of occurrence in the last 10 years is 45\%. In this
setting an expert could always assign a constant probability of 0.45
and be well-calibrated. This assessment, however, can be made without
any subject-matter expertise. For this reason the long-term frequency
is often considered the baseline probability---a naive assessment that
provides the decision-maker very little extra information. Experts
should make probability assessments that are as far from the baseline
as possible. The extent to which their probabilities differ from the
baseline is measured by \textit{sharpness} [\citet
{gneiting2008rejoinder,winkler2008comments}]. If the experts are both
sharp and well calibrated, they can forecast the behavior of the
process with high certainty and accuracy. Therefore, useful probability
estimation should maximize sharpness subject to calibration [see, e.g.,
\citet{raftery2005using,murphy1987general}].

There is strong empirical evidence that bringing together the strengths
of different experts by combining their probability forecasts into a
single consensus, known as the \textit{crowd belief}, improves
predictive performance. Prompted by the many applications of
probability forecasts, including medical diagnosis [\citet
{wilson1998prediction,pepe2003statistical}], political and
socio-economic foresight [\citet{tetlock2005expert}], and meteorology
[\citet{sanders1963subjective,vislocky1995improved,baars2005performance}], researchers have proposed many approaches to
combining probability forecasts
[see, e.g., \citet{Ranjan08,satopaa,batchelder2010cultural} for some
recent studies, and \citet{Genest,Wallsten97evaluatingand,clemen2007aggregating,primo2009calibration} for a comprehensive
overview]. The general focus, however, has been on developing one-time
aggregation procedures that consult the experts' advice only once
before the event resolves.

Consequently, many areas of probability aggregation still remain rather
unexplored. For instance, consider investors aiming to assess whether a
stock index will finish trading above a threshold on a given date. To
maximize their overall predictive accuracy, they may consult a group of
experts repeatedly over a period of time and adjust their estimate of
the aggregate probability accordingly. Given that the experts are
allowed to update their probability assessments, the aggregation should
be performed by taking into account the temporal correlation in their advice.

This paper adds another layer of complexity by assuming a heterogeneous
set of experts, most of whom only make one or two probability
assessments over the hundred or so days before the event resolves. This
means that the decision-maker faces a different group of experts every
day, with only a few experts returning later on for a second round of
advice. The problem at hand is therefore strikingly different from many
time-series estimation problems, where one has an observation at every
time point---or almost every time point. As a result, standard
time-series procedures like ARIMA [see, e.g., \citet{mills1991time}]
are not directly applicable. This paper introduces a time-series model
that incorporates self-reported expertise and captures a sharp and
well-calibrated estimate of the crowd belief. The model is highly
interpretable and can be used for the following:
\begin{itemize}
\item analyzing under and overconfidence in different groups of experts,
\item obtaining accurate probability forecasts, and
\item gaining question-specific quantities with easy interpretations,
such as expert disagreement and problem difficulty.
\end{itemize}

This paper begins by describing our geopolitical database. It then
introduces a dynamic hierarchical model for capturing the crowd belief.
The model is estimated in a two-step procedure: first, a sampling step
produces constrained parameter estimates via Gibbs sampling [see, e.g.,
\citet{geman1984stochastic}]; second, a calibration step transforms
these estimates to their unconstrained equivalents via a
one-dimensional optimization procedure. The model introduction is
followed by the first evaluation section that uses synthetic data to
study how accurately the two-step procedure can estimate the crowd
belief. The second evaluation section applies the model to our
real-world geopolitical forecasting database. The paper concludes with
a discussion of future research directions and model limitations.

\section{Geopolitical forecasting data}
\label{data}
Forecasters were recruited from professional societies, research
centers, alumni associations, science bloggers and word of mouth ($n =
2365$). Requirements included at least a Bachelor's degree and
completion of psychological and political tests that took roughly two
hours. These measures assessed cognitive styles, cognitive abilities,
personality traits, political attitudes and real-world knowledge. The
experts were asked to give probability forecasts (to the second decimal
point) and to self-assess their level of expertise (on a 1-to-5 scale
with 1${} = {}$Not At All Expert and 5${} = {}$Extremely Expert) on a
number of 166
geopolitical binary events taking place between September 29, 2011 and
May 8, 2013. Each question was active for a period during which the
participating experts could update their forecasts as frequently as
they liked without penalty. The experts knew that their probability
estimates would be assessed for accuracy using Brier scores.\setcounter{footnote}{1}\footnote{The Brier score is the squared distance between the probability
forecast and the event indicator that equals 1.0 or 0.0 depending on
whether the event happened or not, respectively. See \citet{Brier} for
the original introduction.} This incentivized them to report their
true beliefs instead of attempting to game the system [\citet
{winkler1968good}]. In addition to receiving \$150 for meeting minimum
participation requirements that did not depend on prediction accuracy,
the experts received status rewards for their performance via
leader-boards displaying Brier scores for the top 20 experts. Given
that a typical expert participated only in a small subset of the 166
questions, the experts are considered indistinguishable conditional on
the level of self-reported expertise.

The average number of forecasts made by a single expert in one day was
around 0.017, and the average group-level response rate was around 13.5
forecasts per day. Given that the group of experts is large and
diverse, the resulting data set is very sparse. Tables~\ref{DataStats}
and \ref{ExpertiseTable} provide relevant summary statistics on the
data. Notice that the distribution of the self-reported expertise is
skewed to the right and that some questions remained active longer than
others. For more details on the data set and its collection see \citet
{ungar2012good}.

%
\begin{table}
\tabcolsep=0pt
\caption{Five-number summaries of our real-world data}
\label{DataStats}
\begin{tabular*}{\textwidth}{@{\extracolsep{\fill}}ld{3.0}d{3.1}d{3.1}d{3.1}
d{3.2}d{4.0}@{}}
\hline
\multicolumn{1}{@{}l}{\textbf{Statistic}} & \multicolumn
{1}{c}{\textbf{Min.}} &
\multicolumn{1}{c}{$\bolds{Q_1}$} & \multicolumn{1}{c}{\textbf
{Median}} &
\multicolumn{1}{c}{\textbf{Mean}} & \multicolumn{1}{c}{$\bolds
{Q_3}$} &
\multicolumn{1}{c}{\textbf{Max.}}\\
\hline
\# of days a question is active & 4 & 35.6 & 72.0 & 106.3 & 145.20 &
418 \\
\# of experts per question & 212 & 543.2 & 693.5 & 783.7 & 983.2 &
1690\\
\# forecasts given by each expert on a question & 1 & 1.0 & 1.0 & 1.8 &
2.0 & 131 \\
\# questions participated by an expert & 1 & 14.0 & 36.0 & 55.0 &
90.0 & 166 \\
\hline
\end{tabular*}
\end{table}

%
\begin{table}[b]
\caption{Frequencies of the self-reported expertise
(1${} = {}$Not At All Expert and 5${} = {}$Extremely Expert)
levels across all the 166 questions in our real-world data}
\label{ExpertiseTable}
\begin{tabular*}{\textwidth}{@{\extracolsep{\fill
}}ld{2.1}d{2.1}d{2.1}d{1.1}d{1.1}@{}}
\hline
Expertise level & 1 & 2 & 3 & 4 & 5\\
Frequency (\%) & 25.3 & 30.7 & 33.6 & 8.2 & 2.1 \\
\hline
\end{tabular*}
\end{table}

%
%
%
%
%

%
\begin{figure}

\includegraphics{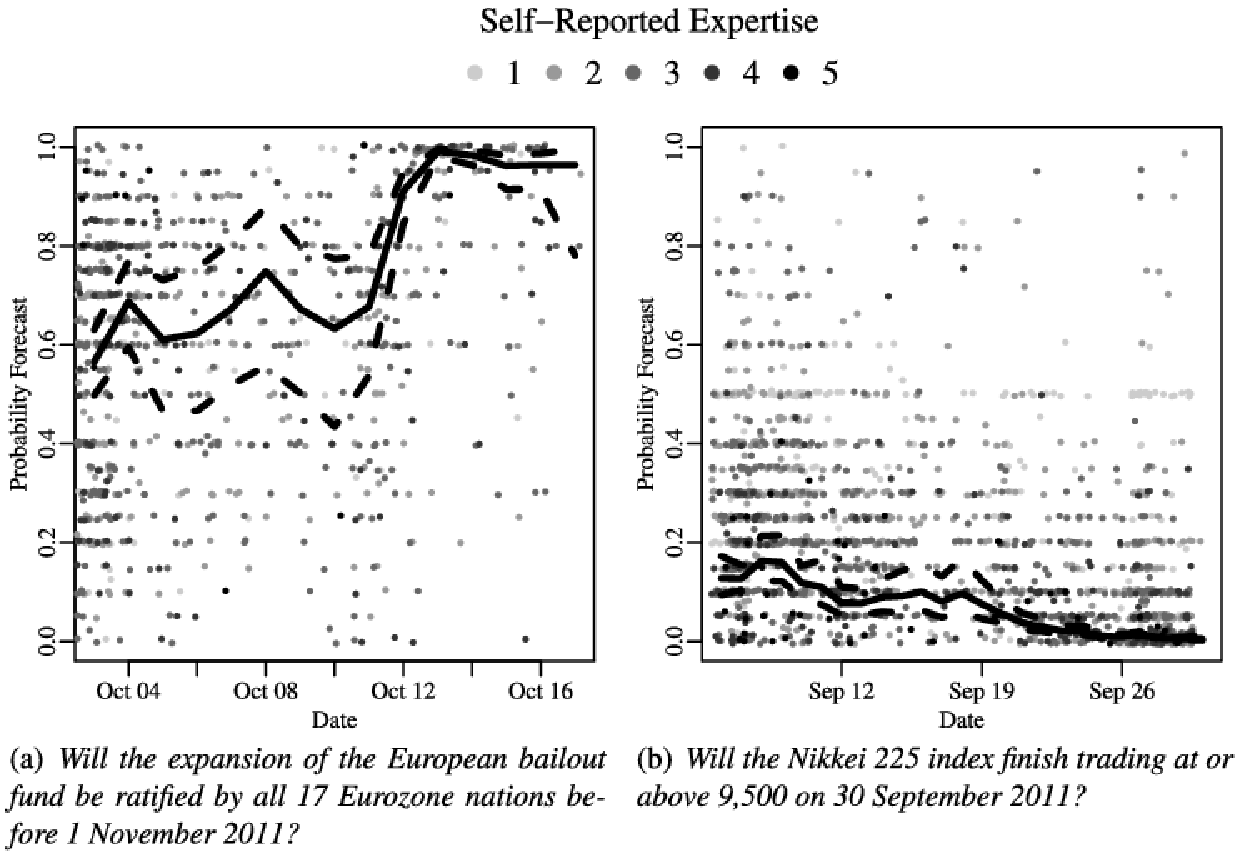}

\caption{Scatterplots of the probability forecasts given for two questions
in our data set. The solid line gives the posterior mean of the
calibrated crowd belief as estimated by our model.
The surrounding dashed lines connect the point-wise 95\% posterior intervals.}
\label{ExamplePlotsFinal}\vspace*{12pt}
\end{figure}
%
%
%
%
%
%

To illustrate the data with some concrete examples,
Figure~\ref{ExamplePlotsFinal}(a)
and \ref{ExamplePlotsFinal}(b) show scatterplots of the probability
forecasts given for (a) \textit{Will the expansion of the European
bailout fund be ratified by all 17 Eurozone nations before 1 November
2011?} and (b) \textit{Will the Nikkei 225 index finish trading at or
above 9500 on 30 September 2011?} The points have been shaded
according to the level of self-reported expertise and jittered slightly
to make overlaps visible. The solid line gives the posterior mean of
the calibrated crowd belief as estimated by our model. The surrounding
dashed lines connect the point-wise 95\% posterior intervals. Given
that the European bailout fund was ratified before November 1, 2011 and
that the Nikkei 225 index finished trading at around 8700 on September
30, 2011, the general trend of the probability forecasts tends to
converge toward the correct answers. The individual experts, however,
sometimes disagree strongly, with the disagreement persisting even near
the closing dates of the questions.

\section{Model}
\label{model}
Let $p_{i,t,k} \in(0,1)$ be the probability forecast given by the
$i$th expert at time $t$ for the $k$th question, where $i = 1, \ldots,
I_k$, $t = 1, \ldots, T_k$, and $k = 1, \ldots, K$. Denote the logit
probabilities with
\[
Y_{i,t,k} = \logit(p_{i,t,k}) = \log \biggl( \frac
{p_{i,t,k}}{1-p_{i,t,k}}
\biggr) \in\R
\]
and collect the logit probabilities for question $k$ at time $t$ into a
vector $\mathbf{Y}_{t,k} = [Y_{1,t,k} Y_{2,t,k} \cdots Y_{I_{k},t,k}]^T$. Partition the experts into $J$ groups based on some
individual feature, such as self-reported expertise, with each group
sharing a common multiplicative bias term $b_{j} \in\R$ for $j = 1,
\ldots, J$. Collect these bias terms into a bias vector $\mathbf
{b} = [b_{1}\ b_{2}\ \cdots\ b_{J}]^T$. Let $\mathbf{M}_k$ be a
$I_k \times J$ matrix denoting the group memberships of the experts in
question $k$; that is, if the $i$th expert participating in the $k$th
question belongs to the $j$th group, then the $i$th row of $\mathbf
{M}_k$ is the $j$th standard basis vector $\mathbf{e}_j$. The bias
vector $\mathbf{b}$ is assumed to be identical across all $K$
questions. Under this notation, the model for the $k$th question can be
expressed as
\begin{eqnarray}
\mathbf{Y}_{t, k} &=& \mathbf{M}_k \mathbf{b}
X_{t, k} + \mathbf{v}_{t, k},\label{observedpr}
\\
X_{t, k} &=& \gamma_k X_{t-1, k} + w_{t, k},
\label{hiddenpr}
\\
X_{0,k} &\sim& \mathcal{N} \bigl(\mu_0,
\sigma^2_0 \bigr),
\nonumber
\end{eqnarray}
where (\ref{observedpr}) denotes the observed process, (\ref
{hiddenpr}) shows the hidden process that is driven by the constant
$\gamma_k \in\R$, and $(\mu_0, \sigma_0^2) \in(\R, \R^+)$ are
hyperparameters fixed {a priori} to 0 and 1, respectively. The
error terms follow:
\begin{eqnarray*}
\mathbf{v}_{t, k} | \sigma^2_k &\stackrel{
\mathrm{i.i.d.}} {\sim}& \mathcal{N}_{I_k} \bigl(\mathbf{0},
\sigma^2_k \mathbf{I}_{I_k} \bigr),
\\
w_{t, k} | \tau^2_k &\stackrel{\mathrm{i.i.d.}}
{\sim}& \mathcal{N} \bigl(0, \tau^2_k \bigr).
\end{eqnarray*}
Therefore, the parameters of the model are $\mathbf{b}$, $\sigma
^2_k$, $\gamma_k$ and $\tau^2_k$ for $k = 1, \ldots, K$. Their prior
distributions are chosen to be noninformative, $p(\mathbf{b},
\sigma^2_k| \mathbf{X}_k) \propto\sigma^2_k$ and $p(\gamma_k,
\tau^2_k| \mathbf{X}_k) \propto\tau^2_k$.

The hidden state $X_{t,k}$ represents the aggregate logit probability
for the $k$th event given all the information available up to and
including time $t$. To make this more specific, let $Z_k \in\{0, 1\}$
indicate whether the event associated with the $k$th question happened
$ (Z_k = 1 )$ or did not happen $ (Z_k = 0 )$. If
$\{\mathcal{F}_{t,k}\}_{t=1}^{T_k}$ is a filtration representing the
information available up to and including a given time point, then
according to our model $\E[Z_k | \mathcal{F}_{t,k}] = \P(Z_k = 1|
\mathcal{F}_{t,k}) = \logit^{-1}(X_{t,k})$. Ideally this probability
maximizes sharpness subject to calibration [for technical definitions
of calibration and sharpness see \citet{Ranjan08,gneiting2013combining}].
%
Even though a single expert is unlikely to have access to all the
available information, a large and diverse group of experts may share a
considerable portion of the available information. The collective
wisdom of the group therefore provides an attractive proxy for
$\mathcal{F}_{t,k}$.

Given that the experts may believe in false information, hide their
true beliefs or be biased for many other reasons, their probability
assessments should be aggregated via a model that can detect potential
bias, separate signal from noise and use the collective opinion to
estimate $X_{t,k}$. In our model the experts are assumed to be, on
average, a multiplicative constant $\mathbf{b}$ away from
$X_{t,k}$. Therefore, an individual element of $\mathbf{b}$ can be
interpreted as a group-specific \textit{systematic bias} that labels
the group either as overconfident [$b_j \in(1, \infty)$] or as
underconfident [$b_j \in(0,1)$]. See Section~\ref{model} for a
brief discussion on different bias structures. Any
other deviation from $X_{t,k}$ is considered \textit{random noise}.
This noise is measured in terms of $\sigma^2_k$ and can be assumed to
be caused by momentary over-optimism (or pessimism), false beliefs or
other misconceptions.

The \textit{random fluctuations} in the hidden process are measured by
$\tau^2_k$ and are assumed to represent changes or shocks to the
underlying circumstances that ultimately decide the outcome of the
event. The \textit{systematic component} $\gamma_k$ allows the model
to incorporate a constant signal stream that drifts the hidden process.
If the uncertainty in the question diminishes [$\gamma_k \in(1,
\infty)$], the hidden process drifts to positive or negative infinity.
Alternatively, the hidden process can drift to zero, in which case any
available information does not improve predictive accuracy [$\gamma_k
\in(0, 1)$]. Given that all the questions in our data set were
resolved within a prespecified timeframe, we expect $\gamma_k \in(1,
\infty)$ for all $k = 1, \ldots, K$.

As for any future time $T^{*}\geq t$,
\begin{eqnarray*}
X_{T^*,k} &=& \gamma_k^{T^*-t}X_t + \sum
_{i=t+1}^{T^*} \gamma _k^{T^*-i}
w_{i}
\\
&\sim& \mathcal{N} \Biggl( \gamma_k^{T^*-t}X_{t,k},
\tau_k^2 \sum_{i=t+1}^{T^*}
\gamma_k^{T^*-i} \Biggr),
\end{eqnarray*}
the model can be used for time-forward prediction as well. The
prediction for the aggregate logit probability at time $T^*$ is given
by an estimate of $\gamma^{T^*-t}X_{t,k}$. Naturally the uncertainty
in this prediction grows in $T$. To make such time-forward predictions,
it is necessary to assume that the past population of experts is
representative of the future population. This is a reasonable
assumption because even though the future population may consist of
entirely different individuals, on average the population is likely to
look very similar to the past population. In practice, however, social
scientists are generally more interested in an estimate of the current
probability than the probability under unknown conditions in the
future. For this reason, our analysis focuses on probability
aggregation only up to the current time $t$.

For the sake of model identifiability, it is sufficient to share only
one of the elements of $\mathbf{b}$ among the $K$ questions. In
this paper, however, all the elements of $\mathbf{b}$ are assumed
to be identical across the questions because some of the questions in
our real-world data set involve very few experts with the highest level
of self-reported expertise. The model can be extended rather easily to
estimate bias at a more general level. For instance, by assuming a
hierarchical structure $b_{ik} \sim\mathcal{N} (b_{j(i,k)},
\sigma^2_{j(i,k)} )$, where $j(i,k)$ denotes the self-reported
expertise of the $i$th expert in question $k$, the bias can be
estimated at an individual-level. These estimates can then be compared
across questions. Individual-level analysis was not performed in our
analysis for two reasons. First, most experts gave only a single
prediction per problem, which makes accurate bias estimation at the
individual-level very difficult. Second, it is unclear how the
individually estimated bias terms can be validated.

If the future event can take upon $M > 2$ possible outcomes, the hidden
state $X_{t,k}$ is extended to a vector of size $M-1$ and one of the
outcomes, for example, the $M$th one, is chosen as the base case to
ensure that the probabilities will sum to one at any given time point.
Each of the remaining $M-1$ possible outcomes is represented by an
observed process similar to (\ref{observedpr}). Given that this
multinomial extension is equivalent to having $M-1$ independent
binary-outcome models, the estimation and properties of the model are
easily extended to the multi-outcome case. This paper focuses on binary
outcomes because it is the most commonly encountered setting in practice.

\section{Model estimation}
\label{identifiability}
This section introduces a two-step procedure, called \textit
{Sample-And-Calibrate} (SAC), that captures a well-calibrated estimate
of the hidden process without sacrificing the interpretability of our model.

\subsection{Sampling step}
\label{sampling_step}
Given that $ (a \mathbf{b}, X_{t,k}/a, a^2 \tau_k^2 )
\neq (\mathbf{b}, X_{t,k}, \tau_k^2 )$ for any \mbox{$a >
0$} yield the same likelihood for $\mathbf{Y}_{t,k}$, the model as
described by (\ref{observedpr}) and (\ref{hiddenpr}) is not
identifiable. A well-known solution is to choose one of the elements of
$\mathbf{b}$, say, $b_3$, as the reference point and fix $b_3 =
1$. In Section~\ref{syntheticData} we provide a guideline for choosing
the reference point. Denote the constrained version of the model by
\begin{eqnarray*}
\mathbf{Y}_{t, k} &=& \mathbf{M}_k \mathbf{b}
(1) X_{t,
k}(1)+ \mathbf{v}_{t, k},
\\
X_{t, k}(1) &=& \gamma_k(1) X_{t-1, k}(1) +
w_{t, k},
\\
\mathbf{v}_{t, k} | \sigma^2_k(1) &
\stackrel{\mathrm{i.i.d.}} {\sim}& \mathcal{N}_{I_k} \bigl(
\mathbf{0}, \sigma^2_k(1) \mathbf
{I}_{I_k} \bigr),
\\
w_{t, k} | \tau^2_k(1) &\stackrel{
\mathrm{i.i.d.}} {\sim}& \mathcal{N} \bigl(0, \tau^2_k(1)
\bigr),
\end{eqnarray*}
where the trailing input notation, (a), signifies the value under the
constraint $b_{3} = a$. Given that this version is identifiable,
estimates of the model parameters can be obtained. Denote the estimates
by placing a hat on the parameter symbol. For instance, $\hat
{\mathbf{b}}(1)$ and $\hat{X}_{t, k}(1)$ represent the estimates
of $\mathbf{b}(1)$ and $X_{t, k}(1)$, respectively.

These estimates are obtained by first computing a posterior sample via
Gibbs sampling and then taking the average of the posterior sample. The
first step of our Gibbs sampler is to sample the hidden states via the
\textit{Forward-Filtering-Backward-Sampling} (FFBS) algorithm. FFBS
first predicts the hidden states using a Kalman filter and then
performs a backward sampling procedure that treats these predicted
states as additional observations [see, e.g., \citet{carter1994gibbs,migon2005dynamic} for details on FFBS]. Given that the Kalman filter
can handle varying numbers or even no forecasts at different time
points, it plays a very crucial role in our probability aggregation
under sparse data.
%
%

Our implementation of the sampling step is written in C$++$ and runs
quite quickly. To obtain 1000 posterior samples for 50 questions each
with 100 time points and 50 experts takes about 215 seconds on a 1.7
GHz Intel Core i5 computer. See the supplemental article for the
technical details of the sampling steps [\citet{satopaa2014}] and, for
example, \citet{gelman2003bayesian} for a discussion on the general
principles of Gibbs sampling.

\subsection{Calibration step}
\label{calibration_step}
Given that the model parameters can be estimated by fixing $b_3$ to any
constant, the next step is to search for the constant that gives an
optimally sharp and calibrated estimate of the hidden process. This
section introduces an efficient procedure that finds the optimal
constant without requiring any additional runs of the sampling step.
First, assume that parameter estimates $\hat{\mathbf{b}}(1)$ and
$\hat{X}_{t, k}(1)$ have already been obtained via the sampling step
described in Section~\ref{sampling_step}. Given that for any $\beta
\in\R/ \{0\}$,
\begin{eqnarray*}
\mathbf{Y}_{t, k} &=& \mathbf{M}_k \mathbf{b}
(1) X_{t,
k}(1)+ \mathbf{v}_{t, k}
\\
&=& \mathbf{M}_k \bigl(\mathbf{b} (1) \beta \bigr)
\bigl(X_{t, k}(1)/\beta \bigr) + \mathbf{v}_{t, k}
\\
&=& \mathbf{M}_k \mathbf{b} (\beta) X_{t, k}(\beta)
+ \mathbf{v}_{t, k},
\end{eqnarray*}
we have that $\mathbf{b} (\beta) = \mathbf{b} (1) \beta$
and $X_{t, k}(\beta) = X_{t, k}(1)/\beta$.
Recall that the hidden process $X_{t, k}$ is assumed to be sharp and
well calibrated. Therefore, $b_3$ can be estimated with the value of
$\beta$ that simultaneously maximizes the sharpness and calibration of
$\hat{X}_{t, k}(1) / \beta$. A natural criterion for this
maximization is given by the class of \textit{proper scoring rules}
that combine sharpness and calibration [\citet{gneiting2008rejoinder,buja2005loss}]. Due to the possibility of \textit{complete separation}
in any one question [see, e.g., \citet{gelman2008weakly}], the
maximization must be performed over multiple questions. Therefore,
\begin{equation}
\label{OSE} \hat{\beta}= \argmax_{\beta\in\R/ \{0\}} \sum
_{k=1}^K \sum_{t=1}^{T_k}
S \bigl(Z_k, \hat{X}_{k,t}(1) / \beta \bigr),
\end{equation}
where $Z_k \in\{0, 1\}$ is the event indicator for question $k$. The
function $S$ is a strictly proper scoring rule such as the negative
Brier score [\citet{Brier}]
\[
S_{\mathrm{BRI}}(Z, X) = - \bigl(Z - \logit^{-1}(X)
\bigr)^2
\]
or the logarithmic score [\citet{good1952rational}]
\[
S_{\mathrm{LOG}}(Z, X) = Z \log \bigl(\logit^{-1}(X) \bigr) + (1-Z)
\log \bigl(1-\logit^{-1}(X) \bigr).
\]
%
The estimates of the unconstrained model parameters are then given by
\begin{eqnarray*}
\hat{X}_{t,k}&=& \hat{X}_{k,t}(1) / \hat{\beta},
\\
\hat{\mathbf{b}}&=& \hat{\mathbf{b}}(1) \hat{\beta},
\\
\hat{\tau}_{k}^2&=& \hat{\tau}_{k}^2(1)/
\hat{\beta}^2,
\\
\hat{\sigma}_{k}^2&=& \hat{\sigma}_{k}^2(1),
\\
\hat{\gamma}_{k}&=& \hat{\gamma}_{k}(1).
\end{eqnarray*}
Notice that estimates of $\sigma^2_k$ and $\gamma_k$ are not affected
by the constraint.

%

\section{Synthetic data results}
\label{syntheticData}

This section uses synthetic data to evaluate how accurately the
SAC-procedure captures the hidden states and bias vector.
The hidden process is generated from standard Brownian motion. More
specifically, if $Z_{t,k}$ denotes the value of a path at time $t$, then
\begin{eqnarray*}
Z_k &=& \mathbbm{1} ( Z_{T_k, k} > 0 ),
\\
X_{t,k} &=& \logit \biggl[ \Phi \biggl(\frac{Z_{t, k}}{\sqrt{T_k- t}} \biggr)
\biggr]
\end{eqnarray*}
gives a sequence of $T_k$ calibrated logit probabilities for the event
$Z_k = 1$. A hidden process is generated for $K$ questions with a time
horizon of $T_k = 101$. The questions involve 50 experts allocated
evenly among five expertise groups. Each expert gives one probability
forecast per day with the exception of time $t = 101$ when the event
resolves. The forecasts are generated by applying bias and noise to the
hidden process as described by (\ref{observedpr}). Our simulation
study considers a three-dimensional grid of parameter values:
\begin{eqnarray*}
\sigma^2 &\in& \{1/2, 1, 3/2, 2, 5/2\},
\\
\beta&\in& \{1/2, 3/4, 1, 4/3, 2/1\},
\\
K &\in& \{20, 40, 60, 80, 100\},
\end{eqnarray*}
where $\beta$ varies the bias vector by $\mathbf{b} = [1/2, 3/4,
1, 4/3, 2/1]^T \beta$. Forty synthetic data sets are generated for
each combination of $\sigma^2$, $\beta$ and $K$ values. The
SAC-procedure runs for 200 iterations of which the first 100 are used
for burn-in.

%

%
%


%
%
%
%
%
%
%
%
%
%


SAC under the Brier ($\mbox{SAC}_{\mathrm{BRI}}$) and logarithm score
($\mbox{SAC}_{\mathrm{LOG}}$) are compared with the \textit
{Exponentially Weighted Moving Average} (EWMA). EWMA, which serves as a
baseline, can be understood by first denoting the (expertise-weighted)
average forecast at time $t$ for the $k$th question with
\begin{equation}
\label{weighted mean} \bar{p}_{t,k} = \sum_{j=1}^J
\omega_j \biggl( \frac{1}{|E_j|}\sum_{i \in E_j}
p_{i,t,k} \biggr),
\end{equation}
where $E_j$ refers to an index set of all experts in the $j$th
expertise group and $\omega_j$ denotes the weight associated with the
$j$th expertise group. The EWMA forecasts for the $k$th problem are
then constructed recursively from
\[
\hat{p}_{t,k}(\alpha) = \cases{ \bar{p}_{1,k}, &\quad $\mbox{for }
t = 1$, \vspace*{2pt}
\cr
\alpha\bar{p}_{t,k} + (1-\alpha)
\hat{p}_{t-1,k}(\alpha), & \quad$\mbox {for } t > 1$,}
\]
where $\alpha$ and $\bolds{\omega}$ are learned from the
training set by
\[ 
(\hat{\alpha}, \hat{\bolds{\omega}}) = \argmin_{\alpha,
\omega_j \in[0,1]} \sum
_{k =1}^K \sum_{t=1}^{T_k}
\bigl(Z_k - \hat {p}_{t,k}(\alpha, \bolds{\omega})
\bigr)^2  \qquad\mbox{s.t. } \sum
_{j=1}^J \omega_j = 1.
\]

%
If $p_{t,k} = \logit^{-1}(X_{t,k})$ and $\hat{p}_{t,k}$ is the
corresponding probability estimated by the model, the model's accuracy
to estimate the hidden process is measured with the quadratic loss,
$(p_{t,k} - \hat{p}_{t,k})^2$, and the absolute loss, $|p_{t,k} - \hat
{p}_{t,k}|$. Table~\ref{lossesSynth} reports these losses averaged
over all conditions, simulations and time points.
The three competing methods, $\mbox{SAC}_{\mathrm{BRI}}$, $\mbox
{SAC}_{\mathrm{LOG}}$ and EWMA, estimate the hidden process with great
accuracy.
%
Based on other performance measures that are not shown for the sake of
brevity, all three methods suffer from an increasing level of noise in
the expert logit probabilities but can make efficient use of extra data.

%
\begin{table}
\tablewidth=250pt
\caption{Summary measures of the estimation accuracy under synthetic data.
As EWMA does not produce an estimate of the bias vector,
its accuracy on the bias term cannot be reported}\label{lossesSynth}
\begin{tabular*}{250pt}{@{\extracolsep{\fill}}lcc@{}}
\hline
\textbf{Model} & \textbf{Quadratic loss} & \textbf{Absolute loss}\\
\hline
\multicolumn{3}{c}{Hidden process}\\
$\mbox{SAC}_{\mathrm{BRI}}$ & 0.00226 & 0.0334 \\
$\mbox{SAC}_{\mathrm{LOG}}$ & 0.00200 & 0.0313 \\
EWMA & 0.00225 & 0.0339\\[3pt]
\multicolumn{3}{c}{Bias vector}\\
$\mbox{SAC}_{\mathrm{BRI}}$ & 0.147\phantom{00} & 0.217\phantom{0}
\\
$\mbox{SAC}_{\mathrm{LOG}}$ & 0.077\phantom{00} & 0.171\phantom{0}
\\
\hline
\end{tabular*}
\end{table}
%

%
\begin{figure}[b]

\includegraphics{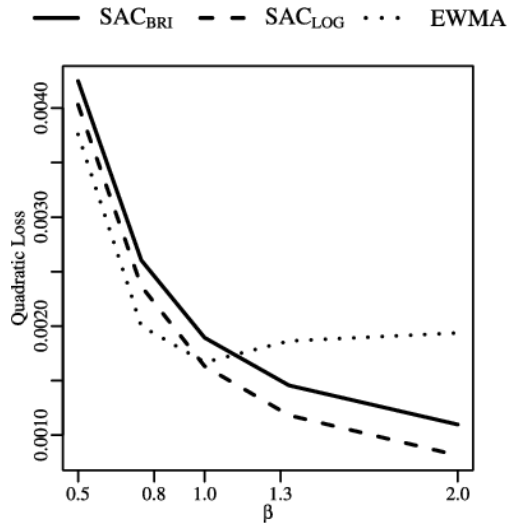}

\caption{The marginal effect of $\beta$ on the average quadratic loss.}
\label{Synthetic}
\end{figure}


%
%
Some interesting differences emerge from Figure~\ref{Synthetic} which
shows the marginal effect of $\beta$ on the average quadratic loss. As
can be expected, EWMA performs well when the experts are, on average,
close to unbiased. Interestingly, SAC estimates the hidden process more
accurately when the experts are overconfident (large $\beta$) compared
to underconfident (small $\beta$). To understand this result, assume
that the experts in the third group are highly underconfident. Their
logit probabilities are then expected to be closer to zero than the
corresponding hidden states. After adding white noise to these expected
logit probabilities, they are likely to cross to the other side of
zero. If the sampling step fixes $b_3 = 1$, as it does in our case, the
third group is treated as unbiased and some of the constrained
estimates of the hidden states are likely to be on the other side of
zero as well. Unfortunately, this discrepancy cannot be corrected by
the calibration step that is restricted to shifting the constrained
estimates either closer or further away from zero but not across it. To
maximize the likelihood of having all the constrained estimates on the
right side of zero and hence avoiding the discrepancy, the reference
point in the sampling step should be chosen with care. A helpful
guideline is to fix the element of $\mathbf{b}$ that is {a
priori} believed to be the largest.

The accuracy of the estimated bias vector is measured with the
quadratic loss, $(b_j - \hat{b}_j)^2$, and the absolute loss, $|b_j -
\hat{b}_j|$. Table~\ref{lossesSynth} reports these losses averaged
over all conditions, simulations and elements of the bias vector.
Unfortunately, EWMA does not produce an estimate of the bias vector.
Therefore, it cannot be used as a baseline for the estimation accuracy
in this case. Given that the losses for $\mbox{SAC}_{\mathrm{BRI}}$
and $\mbox{SAC}_{\mathrm{LOG}}$ are quite small, they estimate the
bias vector accurately.

\section{Geopolitical data results}
\label{realData}

This section presents results for the real-world data described in
Section~\ref{data}. The goal is to provide application specific
insight by discussing the specific research objectives itemized in
Section~\ref{intro}. First, however, we discuss two practical matters
that must be taken into account when aggregating real-world probability
forecasts.

\subsection{Incoherent and imbalanced data}
\label{practicalMatters}
The first matter regards human experts making probability forecasts of
0.0 or 1.0 even if they are not completely sure of the outcome of the
event. For instance, all 166 questions in our data set contain both a
zero and a one. Transforming such forecasts into the logit space yields
infinities that can cause problems in model estimation. To avoid this,
\citet{Ariely00theeffects} suggest changing $p = 0.00$ and $1.00$ to
$p = 0.02$ and $0.98$, respectively. This is similar to \textit
{winsorising} that sets the extreme probabilities to a specified
percentile of the data [see, e.g., \citet{hastings1947low} for more
details on winsorising]. \citet{Geo}, on the other hand, consider only
probabilities that fall within a constrained interval, say, $[0.001,
0.999]$, and discard the rest. Given that this implies ignoring a
portion of the data, we adopt a censoring approach similar to \citet
{Ariely00theeffects} by changing $p = 0.00$ and $1.00$ to $p = 0.01$
and $0.99$, respectively.\vadjust{\goodbreak} Our results remain insensitive to the exact
choice of censoring as long as this is done in a reasonable manner to
keep the extreme probabilities from becoming highly influential in the
logit space.

The second matter is related to the distribution of the class labels in
the data. If the set of occurrences is much larger than the set of
nonoccurrences (or {vice versa}), the data set is called
\textit{imbalanced}. On such data the modeling procedure can end up
over-focusing on the larger class and, as a result, give very accurate
forecast performance over the larger class at the cost of performing
poorly over the smaller class [see, e.g., \citet{chen2009learning,wallace2012class}]. Fortunately, it is often possible to use a
well-balanced version of the data. The first step is to find a
partition $S_0$ and $S_1$ of the question indices $\{1, 2, \ldots, K\}$
such that the equality $\sum_{k \in S_0} T_k = \sum_{k \in S_1} T_k$
is as closely approximated as possible. This is equivalent to an
NP-hard problem known in computer science as the \textit{Partition
Problem}: determine whether a given set of positive integers can be
partitioned into two sets such that the sums of the two sets are equal
to each other [see, e.g., \citet{karmarkar1982differencing,hayes2002easiest}]. A simple solution is to use a greedy algorithm that
iterates through the values of $T_k$ in descending order, assigning
each $T_k$ to the subset that currently has the smaller sum [see, e.g.,
\citet{kellerer2004knapsack,gent1996phase} for more details on the
\textit{Partition Problem}]. After finding a well-balanced partition,
the next step is to assign the class labels such that the labels for
the questions in $S_x$ are equal to $x$ for $x = 0$ or $1$. Recall from
Section~\ref{calibration_step} that $Z_k$ represents the event
indicator for the $k$th question. To define a balanced set of
indicators $ \tilde{Z}_k$ for all $k \in S_x$, let
\begin{eqnarray*}
\tilde{Z}_k &=& x,
\\
\tilde{p}_{i,t,k} &=& \cases{ 1-p_{i,t,k}, &\quad $\mbox{if }
Z_k = 1-x$,\vspace*{2pt}
\cr
p_{i,t,k}, & \quad$\mbox{if }
Z_k = x$,}
\end{eqnarray*}
where $i = 1, \ldots, I_k$, and $t = 1, \ldots, T_k$. The resulting set
\[
\bigl\{ \bigl(\tilde{Z}_k, \{\tilde{p}_{i,t,k} | i = 1,
\ldots, I_k, t = 1, \ldots, T_k \} \bigr) \bigr
\}_{k=1}^K
\]
is a balanced version of the data.
This procedure was used to balance our real-world data set both in
terms of events and time points. The final output splits the events
exactly in half ($|S_0| = |S_1| = 83$) such that the number of time
points in the first and second halves are 8737 and 8738, respectively.

\subsection{Out-of-sample aggregation}
\label{forecasting}
The goal of this section is to evaluate the accuracy of the aggregate
probabilities made by SAC and several other procedures. The models are
allowed to utilize a training set before making aggregations on an
independent testing set. To clarify some of the upcoming notation, let
$S_{\mathrm{train}}$ and $S_{\mathrm{test}}$ be index sets that partition the data into
training and testing sets of sizes $|S_{\mathrm{train}}| = N_{\mathrm{train}}$ and
$|S_{\mathrm{test}}| = 166 - N_{\mathrm{train}}$, respectively.\vadjust{\goodbreak} This means that the $k$th
question is in the training set if and only if $k \in S_{\mathrm{train}}$.
Before introducing the competing models, note that all choices of
thinning and burn-in made in this section are conservative and have
been made based on pilot runs of the models. This was done to ensure a
posterior sample that has low autocorrelation and arises from a
converged chain. The competing models are as follows:

\begin{enumerate}
\item\textit{Simple Dynamic Linear Model} (SDLM). This is equivalent
to the dynamic model from Section~\ref{model} but with $\mathbf
{b} = \mathbf{1}$ and $\beta= 1$. Thus,
\begin{eqnarray*}
\mathbf{Y}_{t, k} &=& X_{t, k} + \mathbf{v}_{t, k},
\\
X_{t, k} &=& \gamma_k X_{t-1, k} + w_{t, k},
\end{eqnarray*}
where $X_{t,k}$ is the aggregate logit probability. Given that this
model does not share any parameters across questions, estimates of the
hidden process can be obtained directly for the questions in the
testing set without fitting the model first on the training set. The
Gibbs sampler is run for 500 iterations of which the first 200 are used
for burn-in. The remaining 300 iterations are thinned by discarding
every other observation, leaving a final posterior sample of 150
observations. The average of this sample gives the final estimates.

\item\textit{The Sample-And-Calibrate procedure both under the Brier}
($\mbox{SAC}_{\mathrm{BRI}}$) \textit{and the Logarithmic score} ($\mbox
{SAC}_{\mathrm{LOG}}$). The model is first fit on the training set by
running the sampling step for 3000 iterations of which the first 500
iterations are used for burn-in. The remaining 2500 observations are
thinned by keeping every fifth observation. The calibration step is
performed for the final 500 observations. The out-of-sample aggregation
is done by running the sampling step for 500 iterations with each
consecutive iteration reading in and conditioning on the next value of
$\beta$ and $\mathbf{b}$ found during the training period. The
first 200 iterations are used for burn-in. The remaining 300 iterations
are thinned by discarding every other observation, leaving a final
posterior sample of 150 observations. The average of this sample gives
the final estimates.

\item\textit{A fully Bayesian version of} $\mbox{SAC}_{\mathrm
{LOG}}$ ($\mbox{BSAC}_{\mathrm{LOG}}$). Denote the calibrated logit
probabilities and event indicators across all $K$ questions with
$\mathbf{X}(1)$ and $\mathbf{Z}$, respectively. The posterior
distribution of $\beta$ conditional on $\mathbf{X}(1)$ is given
by $p(\beta| \mathbf{X}(1), \mathbf{Z}) \propto p(
\mathbf{Z} | \beta, \mathbf{X}(1)) p(\beta| \mathbf
{X}(1))$. The likelihood is
\begin{eqnarray}\label{OSE2}
&& p \bigl( \mathbf{Z} | \beta, \mathbf{X}(1) \bigr)
\nonumber
\\[-8pt]
\\[-8pt]
\nonumber
&&\qquad \propto\prod
_{k=1}^K \prod_{t=1}^{T_k}
\logit^{-1} \bigl(X_{t,k}(1)/\beta \bigr)^{Z_k}
\bigl( 1- \logit^{-1} \bigl( X_{t,k}(1)/\beta \bigr)
\bigr)^{1-Z_k}.
\end{eqnarray}
As in \citet{gelman2003bayesian}, the prior for $\beta$ is chosen to
be locally uniform, $p(1/\beta) \propto1$. Given that this model
estimates $X_{t,k}(1)$ and $\beta$ simultaneously, it is a little
more\vadjust{\goodbreak}
flexible than $\mbox{SAC}$. Posterior estimates of $\beta$ can be
sampled from (\ref{OSE2}) using generic sampling algorithms such as
the Metropolis algorithm [\citet{metropolis1953equation}] or slice
sampling [\citet{neal2003slice}]. Given that the sampling procedure
conditions on the event indicators, the full conditional distribution
of the hidden states is not in a standard form. Therefore, the
Metropolis algorithm is also used for sampling the hidden states.
Estimation is made with the same choices of thinning and burn-in as
described under \textit{Sample-And-Calibrate}.

\item Due to the lack of previous literature on dynamic aggregation of
expert probability forecasts, the main competitors are exponentially
weighted versions of procedures that have been proposed for static
probability aggregation:

\begin{enumerate}[(a)]
\item[(a)]\textit{Exponentially Weighted Moving Average} (EWMA) as
described in Section~\ref{syntheticData}.

\item[(b)] \textit{Exponentially Weighted Moving Logit Aggregator}
(EWMLA). This is a moving version of the aggregator $\hat
{p}_G(\mathbf{b})$ that was introduced in \citet{satopaa}.
%
The EWMLA aggregate probabilities are found recursively from
\[
\hat{p}_{t,k}(\alpha, \mathbf{b}) = \cases{ G_{1,k}(
\mathbf{b} ), & \quad$\mbox{for } t = 1$, \vspace*{2pt}
\cr
\alpha
G_{t,k}(\mathbf{b} ) + (1-\alpha) \hat{p}_{t-1,k}(\alpha,
\mathbf{b}), &\quad $\mbox{for } t > 1$,}
\]
where the vector $\mathbf{b} \in\mathbb{R}^J$ collects the bias
terms of the expertise groups, and
\fontsize{10pt}{\baselineskip}\selectfont
\[
G_{t,k}(\nu) = \Biggl( \prod_{i=1}^{N_{t,k}}
\biggl( \frac{p_{i, t,
k}}{1-p_{i, t, k}} \biggr)^{ {b_{j(i,k)}}/{N_{t,k}}} \Biggr) \bigg/
\Biggl(1+ \prod
_{i=1}^{N_{t,k}} \biggl( \frac{p_{i, t,
k}}{1-p_{i, t, k}}
\biggr)^{{ b_{j(i,k)}}/{N_{t,k}}} \Biggr).
\]
\normalsize
%
The parameters $\alpha$ and $\mathbf{b}$ are learned from the
training set by
\[
(\hat{\alpha}, \hat{ \mathbf{b}}) = \argmin_{ \mathbf
{b} \in\R^5, \alpha\in[0,1]} \sum
_{k \in S_{\mathrm{train}}} \sum_{t=1}^{T_k}
\bigl(Z_k - \hat{p}_{t,k}(\alpha, \mathbf {b})
\bigr)^2.
\]

\item[(c)] \textit{Exponentially Weighted Moving Beta-transformed
Aggregator\break (EWMBA)}. The static version of the Beta-transformed
aggregator was introduced in \citet{Ranjan08}. A dynamic version can be
obtained by replacing $G_{t,k}(\nu)$ in the EWMLA description with
$H_{\nu, \tau}  ( \bar{p}_{t,k} )$, where $H_{\nu, \tau
}$ is the cumulative distribution function of the Beta distribution and
$\bar{p}_{t,k}$ is given by (\ref{weighted mean}). The parameters
$\alpha, \nu, \tau$ and $\bolds{\omega}$ are learned from the
training set by
\begin{eqnarray}
(\hat{\alpha}, \hat{\nu}, \hat{\tau}, \hat{\bolds{\omega }}) =
\argmin_{\nu, \tau> 0\ \alpha, \omega_j \in[0,1]} \sum_{k \in S_{\mathrm{train}}} \sum
_{t=1}^{T_k} \bigl(Z_k -
\hat{p}_{t,k}(\alpha, \nu, \tau, \bolds{\omega})
\bigr)^2
\nonumber
\\[-8pt]
\\[-8pt]
\eqntext{\displaystyle\mbox{s.t. } \sum_{j=1}^J
\omega_j = 1.}
\end{eqnarray}
\end{enumerate}
\end{enumerate}

The competing models are evaluated via a 10-fold
cross-validation\footnote{A 5-fold cross-validation was also
performed. The results were, however, very similar to the 10-fold
cross-validation and hence not presented in the paper.} that first
partitions the 166 questions into 10 sets such that each set has
approximately the same number of questions (16 or 17 questions in our
case) and the same number of time points (between 1760 and 1764 time
points in our case). The evaluation then iterates 10 times, each time
using one of the 10 sets as the testing set and the remaining 9 sets as
the training set. Therefore, each question is used nine times for
training and exactly once for testing. The testing proceeds
sequentially one testing question at a time as follows: First, for a
question with a time horizon of $T_k$, give an aggregate probability at
time $t=2$ based on the first two days. Compute the Brier score for
this probability. Next give an aggregate probability at time $t=3$
based on the first three days and compute the Brier score for this
probability. Repeat this process for all of the $T_k-1$ days. This
leads to $T_k-1$ Brier scores per testing question and a total of
17,475 Brier scores across the entire data set.

%
\begin{table}
\caption{Brier scores based on 10-fold cross-validation.
\textit{Scores by Day} weighs a question by the number of days
the question remained open. \textit{Scores by Problem} gives each
question an equal weight regardless of how long the question remained open.
The bolded values indicate the best scores in each column.
The values in the parenthesis represent standard errors in the
scores}\label{prediction}
\begin{tabular*}{\textwidth}{@{\extracolsep{\fill}}lcccc@{}}
\hline
\textbf{Model} & \multicolumn{1}{c}{\textbf{All}} &
\multicolumn{1}{c}{\textbf{Short}} &
\multicolumn{1}{c}{\textbf{Medium}} &
\multicolumn{1}{c}{\textbf{Long}}\\
\hline
\multicolumn{5}{c}{Scores by day}\\
SDLM & 0.100 (0.156) & 0.066 (0.116) & 0.098 (0.154) & 0.102 (0.157)\\
$\mbox{BSAC}_{\mathrm{LOG}}$ & 0.097 (0.213) & \textbf{0.053}
(0.147) & 0.100 (0.215) & 0.098 (0.215)\\
$\mbox{SAC}_{\mathrm{BRI}}$ & 0.096 (0.190) & 0.056 (0.134) & 0.097
(0.190) & 0.098 (0.192)\\
$\mbox{SAC}_{\mathrm{LOG}}$ & \textbf{0.096} (0.191) & 0.056 (0.134)
& \textbf{0.096} (0.189) & \textbf{0.098} (0.193)\\
EWMBA & 0.104 (0.204) & 0.057 (0.120) & 0.113 (0.205) & 0.105 (0.206)\\
EWMLA & 0.102 (0.199) & 0.061 (0.130) & 0.111 (0.214) & 0.103 (0.200)\\
EWMA & 0.111 (0.146) & 0.080 (0.101) & 0.116 (0.152) & 0.112 (0.146)\\[3pt]
\multicolumn{5}{c}{Scores by problem}\\
SDLM & 0.089 (0.116) & 0.064 (0.085) & 0.106 (0.141) & 0.092 (0.117)\\
$\mbox{BSAC}_{\mathrm{LOG}}$ & 0.083 (0.160) & \textbf{0.052}
(0.103) & 0.110 (0.198) & 0.085 (0.162)\\
$\mbox{SAC}_{\mathrm{BRI}}$& 0.083 (0.142) & 0.055 (0.096) & 0.106
(0.174) & 0.085 (0.144)\\
$\mbox{SAC}_{\mathrm{LOG}}$ & \textbf{0.082} (0.142) & 0.055 (0.096)
& \textbf{0.105} (0.174) & \textbf{0.085} (0.144)\\
EWMBA & 0.091 (0.157) & 0.057 (0.095) & 0.121 (0.187) & 0.093 (0.164)\\
EWMLA & 0.090 (0.159) & 0.064 (0.109) & 0.120 (0.200) & 0.090 (0.159)\\
EWMA & 0.102 (0.108) & 0.080 (0.075) & 0.123 (0.130) & 0.103 (0.110)\\
\hline
\end{tabular*}
\end{table}

Table~\ref{prediction} summarizes these scores in different ways. The
first option, denoted by \textit{Scores by Day}, weighs each question
by the number of days the question remained open. This is performed by
computing the average of the 17,475 scores. The second option, denoted
by \textit{Scores by Problem}, gives each question an equal weight
regardless of how long the question remained open. This is done by
first averaging the scores within a question and then averaging the
average scores across all the questions. Both scores can be further
broken down into subcategories by considering the length of the
questions. The final three columns of Table~\ref{prediction} divide
the questions into \textit{Short} questions (30 days or fewer),
\textit{Medium} questions (between 31 and 59 days) and \textit{Long}
Problems (60 days or more). The number of questions in these
subcategories were 36, 32 and 98, respectively. The bolded scores
indicate the best score in each column. The values in the parenthesis
quantify the variability in the scores: Under \textit{Scores by Day}
the values give the standard errors of all the scores. Under \textit
{Scores by Problem}, on the other hand, the values represent the
standard errors of the average scores of the different questions.

As can be seen in Table~\ref{prediction}, $\mbox{SAC}_{\mathrm
{LOG}}$ achieves the lowest score across all columns except \textit
{Short} where it is outperformed by $\mbox{BSAC}_{\mathrm{LOG}}$. It
turns out that $\mbox{BSAC}_{\mathrm{LOG}}$ is overconfident (see
Section~\ref{calibration}). This means that $\mbox{BSAC}_{\mathrm
{LOG}}$ underestimates the uncertainty in the events and outputs
aggregate probabilities that are typically too near 0.0 or 1.0.
This results into highly variable performance. The short questions
generally involved very little uncertainty. On such easy questions,
overconfidence can pay off frequently enough to compensate for a few
large losses arising from the overconfident and drastically incorrect forecasts.

SDLM, on the other hand, lacks sharpness and is highly underconfident
(see Section~\ref{calibration}). This behavior is expected, as the
experts are underconfident at the group level (see Section~\ref{ExpertBias}) and SDLM does not use the training set to explicitly
calibrate its aggregate probabilities. Instead, it merely smooths the
forecasts given by the experts. The resulting aggregate probabilities
are therefore necessarily conservative, resulting into high average
scores with low variability.

Similar behavior is exhibited by EWMA that performs the worst of all
the competing models. The other two exponentially weighted aggregators,
EWMLA and EWMBA, make efficient use of the training set and present
moderate forecasting performance in most columns of Table~\ref{prediction}. Neither approach, however, appears to dominate the other.
The high variability and average of their performance scores indicate
that their performance suffers from overconfidence.


\subsection{In- and out-of-sample sharpness and calibration}
\label{calibration}
A calibration plot is a simple tool for visually assessing the
sharpness and calibration of a model.
The idea is to plot the aggregate probabilities against the observed
empirical frequencies. Therefore,
any deviation from the diagonal line suggests poor calibration. A model
is considered underconfident
(or overconfident) if the points follow an S-shaped (or \reflectbox
{S}-shaped) trend. To assess sharpness of the model, it is common
practice to place a histogram of the given forecasts in the corner of
the plot. Given that the data were balanced, any deviation from the the
baseline probability of 0.5 suggests improved sharpness.

%
\begin{figure}

\includegraphics{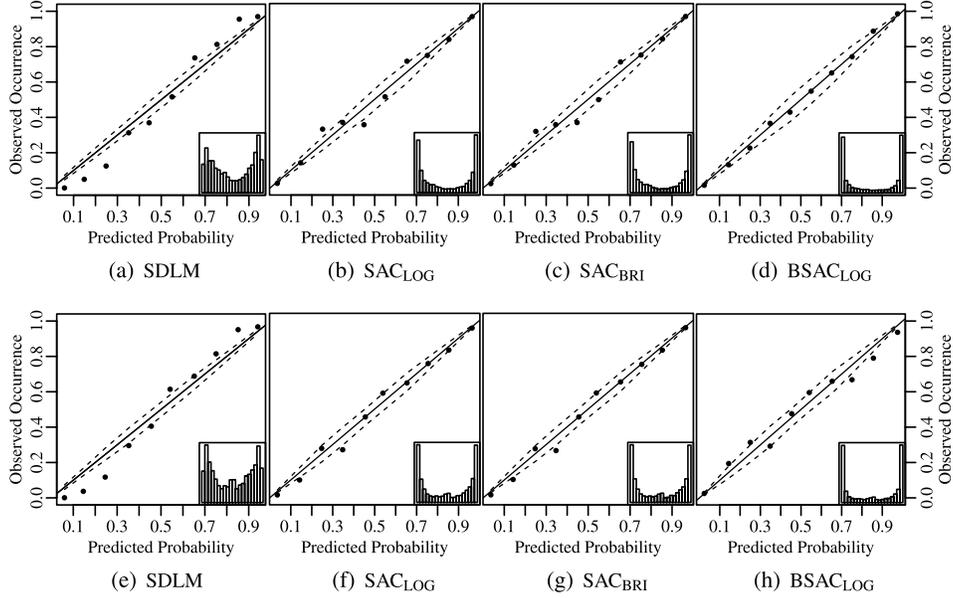}

\caption{The top and bottom rows show in- and out-of-sample calibration
and sharpness, respectively.}\label{Calibration-Out}
\end{figure}

%
%

The top and bottom rows of Figure~\ref{Calibration-Out} present
calibration plots for SDLM, $\mbox{SAC}_{\mathrm{LOG}}$, $\mbox
{SAC}_{\mathrm{BRI}}$ and $\mbox{BSAC}_{\mathrm{LOG}}$ under in- and
out-of-sample probability aggregation, respectively. Each setting is of
interest in its own right: Good in-sample calibration is crucial for
model interpretability. In particular, if the estimated crowd belief is
well calibrated, then the elements of the bias vector $\mathbf{b}$
can be used to study the amount of under or overconfidence in the
different expertise groups. Good out-of-sample calibration and
sharpness, on the other hand, are necessary properties in decision
making. To guide our assessment, the dashed bands around the diagonal
connect the point-wise, Bonferroni-corrected [\citet{bonferroni}] 95\%
lower and upper critical values under the null hypothesis of
calibration. These have been computed by running the bootstrap
technique described in \citet{brocker2007increasing} for 10,000
iterations. The in-sample predictions were obtained by running the
models for 10,200 iterations, leading to a final posterior sample of
1000 observations after thinning and using the first 200 iterations
for burn-in. The out-of-sample predictions were given by the 10-fold
cross-validation discussed in Section~\ref{forecasting}.

Overall, SAC is sharp and well calibrated both in- and out-of-sample
with only a few points barely falling outside the \textit{point-wise}
critical values. Given that the calibration does not change drastically
from the top to the bottom row, SAC can be considered robust against
overfitting. This, however, is not the case with $\mbox{BSAC}_{\mathrm
{LOG}}$ that is well calibrated in-sample but presents overconfidence
out-of-sample. Figure~\ref{Calibration-Out}(a) and (e) serve
as baselines by showing the calibration plots
for SDLM. Given that this model does not perform any explicit
calibration, it is not surprising to see most points outside the
critical values. The pattern in the deviations suggests strong
underconfidence. Furthermore, the inset histogram reveals drastic lack
of sharpness. Therefore, SAC can be viewed as a well-performing
compromise between SDLM and $\mbox{BSAC}_{\mathrm{LOG}}$ that avoids
overconfidence without being too conservative.

\subsection{Group-level expertise bias}
\label{ExpertBias}

%

\begin{figure}

\includegraphics{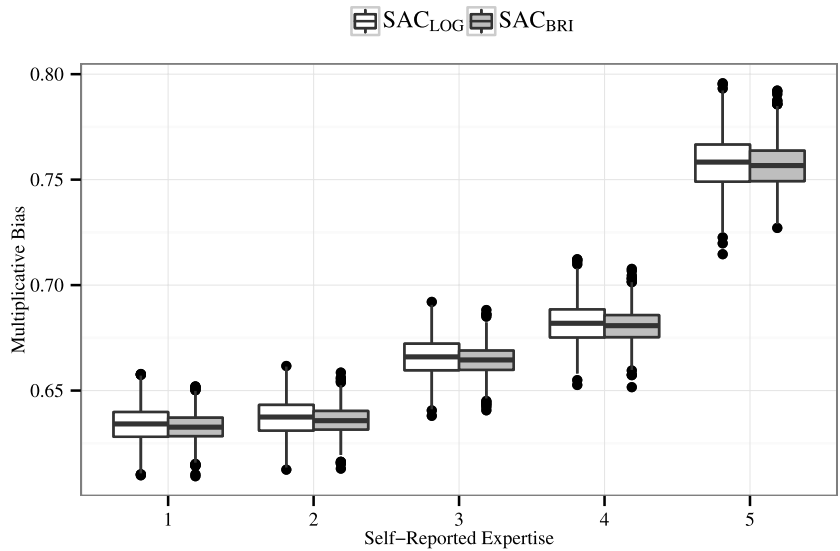}

\caption{Posterior distributions of $b_j$ for $j = 1, \ldots,
5$.}\label{Biases}
\end{figure}

This section explores the bias among the five expertise groups in our
data set. Figure~\ref{Biases} compares the posterior distributions of
the individual elements of $\mathbf{b}$ with side-by-side
boxplots. Given that the distributions fall completely below the
\textit{no-bias} reference line at 1.0, all the expertise groups are
deemed underconfident. Even though the exact level of underconfidence
is affected slightly by the extent to which the extreme probabilities
are censored (see Section~\ref{practicalMatters}), the qualitative
results in this section remain insensitive to different levels of censoring.

Figure~\ref{Biases} shows that underconfidence decreases as expertise
increases. The posterior probability that the most expert group is the
least underconfident is approximately equal to $1.0$, and the posterior
probability of a strictly decreasing level of underconfidence is
approximately 0.87. The latter probability is driven down by the
inseparability of the two groups with the lowest levels of
self-reported expertise. This inseparability suggests that the experts
are poor at assessing how little they know about a topic that is
strange to them. If these groups are combined into a single group, the
posterior probability of a strictly decreasing level of underconfidence
is approximately~1.0.\looseness=1

The decreasing trend in underconfidence can be viewed as a process of
Bayesian updating. A completely ignorant expert aiming to minimize a
reasonable loss function, such as the Brier score, has no reason to
give anything but 0.5 as his probability forecast. However, as soon as
the expert gains some knowledge about the event, he produces an updated
forecast that is a compromise between his initial forecast and the new
information acquired. The updated forecast is therefore conservative
and too close to 0.5 as long as the expert remains only partially
informed about the event. If most experts fall somewhere on this
spectrum between ignorance and full information, their average forecast
tends to fall strictly between 0.5 and the most informed probability
forecast [see \citet{Baron} for more details]. Given that expertise is
to a large extent determined by subject matter knowledge, the level of
underconfidence can be expected to decrease as a function of the
group's level of self-reported expertise.

Finding underconfidence in all the groups may seem like a surprising
result given that many previous studies have shown that experts are
often overconfident [see, e.g.,
\citet{lichtenstein1977calibration,morgan1992uncertainty,bier2004implications} for a summary of numerous
calibration studies]. It is, however, worth emphasizing three points:
First, our result is a statement about groups of experts and hence does
not invalidate the possibility of the individual experts being
overconfident. To make conclusions at the individual level based on the
group level bias terms would be considered an \textit{ecological
inference fallacy} [see, e.g., \citet{lubinski1996seeing}]. Second,
the experts involved in our data set are overall very well calibrated
[\citet{mellers}]. A group of well-calibrated experts, however, can
produce an aggregate forecast that is underconfident. In fact, if the
aggregate is linear, the group is necessarily underconfident [see
Theorem~1 of \citet{Ranjan08}]. Third, according to \citet{Erev1994},
the level of confidence depends on the way the data were analyzed.
%
They explain that experts' probability forecasts suggest
underconfidence when the forecasts are averaged or presented as a
function of independently defined objective probabilities, that is, the
probabilities given by $ \logit^{-1}(X_{t,k})$ in our case. This is
similar to our context and opposite to many empirical studies on
confidence calibration.


\subsection{Question difficulty and other measures}
\label{QuestionDifficulty}
One advantage of our model arises from its ability to produce estimates
of interpretable question-specific parameters $\gamma_k$, $\sigma
^2_k$ and $\tau^2_k$. These quantities can be combined in many
interesting ways to answer questions about different groups of experts
or the questions themselves. For instance, being able to assess the
difficulty of a question could lead to more principled ways of
aggregating performance measures across questions or to novel insight
on the kinds of questions that are found difficult by experts [see,
e.g., a discussion on the \textit{Hard-Easy Effect} in \citet
{Wilson94cognitivefactors}]. To illustrate, recall that higher values
of $\sigma^2_k$ suggest greater disagreement among the participating
experts. Given that experts are more likely to disagree over a
difficult question than an easy one, it is reasonable to assume that
$\sigma^2_k$ has a positive relationship with question difficulty. An
alternative measure is given by $\tau_k^2$ that quantifies the
volatility of the underlying circumstances that ultimately decide the
outcome of the event. Therefore, a high value of $\tau_k^2$ can cause
the outcome of the event to appear unstable and difficult to predict.

As a final illustration of our model, we return to the two example
questions introduced in Figure~\ref{ExamplePlotsFinal}.
Given that $\hat{\sigma}_k^2 = 2.43$ and $\hat{\sigma}_k^2 = 1.77$
for the questions depicted in Figure~\ref{ExamplePlotsFinal}(a) and \ref
{ExamplePlotsFinal}(b), respectively, the first question provokes more
disagreement among the experts than the second one. Intuitively this
makes sense because the target event in Figure~\ref{ExamplePlotsFinal}(a) is
determined by several conditions that may change radically from one day
to the next while the target event in Figure~\ref{ExamplePlotsFinal}(b) is
determined by a relatively steady stock market index. Therefore, it is
not surprising to find that in Figure~\ref{ExamplePlotsFinal}(a) $\hat{\tau
}_k^2 = 0.269$, which is much higher than $\hat{\tau}_k^2 = 0.039$ in
Figure~\ref{ExamplePlotsFinal}(b). We may conclude that the first question is
inherently more difficult than the second one.

\section{Discussion}
This paper began by introducing a rather unorthodox but nonetheless
realistic time-series setting where probability forecasts are made very
infrequently by a heterogeneous group of experts. The resulting data is
too sparse to be modeled well with standard time-series methods. In
response to this lack of appropriate modeling procedures, we propose an
interpretable time-series model that incorporates self-reported
expertise to capture a sharp and well-calibrated estimate of the crowd belief.
This procedure extends the forecasting literature into an
under-explored area of probability aggregation.
%

Our model preserves parsimony while addressing the main challenges in
modeling sparse probability forecasting data. Therefore, it can be
viewed as a basis for many future extensions. To give some ideas,
recall that most of the model parameters were assumed constant over
time. It is intuitively reasonable, however, that these parameters
behave differently during different time intervals of the question. For
instance, the level of disagreement (represented by $\sigma^2_k$ in
our model) among the experts can be expected to decrease toward the
final time point when the question resolves. This hypothesis could be
explored by letting $\sigma^2_{t,k}$ evolve dynamically as a function
of the previous term $\sigma^2_{t-1,k}$ and random noise.

%
This paper modeled the bias separately within each expertise group.
This is by no means restricted to the study of bias or its relation to
self-reported expertise. Different parameter dependencies could be
constructed based on many other expert characteristics, such as gender,
education or specialty, to produce a range of novel insights on the
forecasting behavior of experts. It would also be useful to know how
expert characteristics interact with question types, such as economic,
domestic or international. The results would be of interest to the
decision-maker who could use the information as a basis for hiring only
a high-performing subset of the available experts.



Other future directions could remove some of the obvious limitations of
our model. For instance, recall that the random components are assumed
to follow a normal distribution. This is a strong assumption that may
not always be justified. Logit probabilities, however, have been
modeled with the normal distribution before [see, e.g., \citet
{Erev1994}]. Furthermore, the normal distribution is a rather standard
assumption in psychological models [see, e.g., signal-detection theory
in \citet{tanner1954decision}].

A second limitation resides in the assumption that both the observed
and hidden processes are expected to grow linearly. This assumption
could be relaxed, for instance, by adding higher order terms to the
model. A more complex model, however, is likely to sacrifice
interpretability. Given that our model can detect very intricate
patterns in the crowd belief (see Figure~\ref{ExamplePlotsFinal}),
compromising interpretability for the sake of facilitating nonlinear
growth is hardly necessary.

A third limitation appears in an online setting where new forecasts are
received at a fast rate. Given that our model is fit in a retrospective
fashion, it is necessary to refit the model every time a new forecast
becomes available. Therefore, our model can be applied only to offline
aggregation and online problems that tolerate some delay. A more
scalable and efficient alternative would be to develop an aggregator
that operates recursively on streams of forecasts. Such a \textit
{filtering} perspective would offer an aggregator that estimates the
current crowd belief accurately without having to refit the entire
model each time a new forecast arrives. Unfortunately, this typically
implies being less accurate in estimating the model parameters such as
the bias term. However, as estimation of the model parameters was
addressed in this paper, designing a filter for probability forecasts
seems like the next natural development in time-series probability aggregation.

\section*{Acknowledgments}
The U.S. Government is authorized to reproduce and
distribute reprints for Government purposes notwithstanding any
copyright annotation thereon. Disclaimer: The views and conclusions
expressed herein are those of the authors and should not be
interpreted as necessarily representing the official policies or
endorsements, either expressed or implied, of IARPA, DoI/NBC or the
U.S. Government.\vadjust{\goodbreak}

We deeply appreciate the project management
skills and work of Terry Murray and David Wayrynen, which went far
beyond the call of duty on this project.


\begin{supplement}[id=suppA]
\stitle{Sampling step}
\slink[doi]{10.1214/14-AOAS739SUPP} 
\sdatatype{.pdf}
\sfilename{aoas739\_supp.pdf}
\sdescription{This supplementary material provides a technical description
of the sampling step of the SAC-algorithm.}
\end{supplement}

%

\printaddresses

\end{document}